\documentclass[leqno,11pt,twoside]{article}

\usepackage{amsmath,color}
\usepackage{amsfonts}
\usepackage{pliska}

\def\openone{\leavevmode\hbox{\small1\kern-3.3pt\normalsize1}}

\def\ad{\mbox{ad\,}}

\def\a{{\boldsymbol a}}
\def\b{{\boldsymbol b}}

\def\c{{\boldsymbol c}}
\def\d{{\boldsymbol d}}
\def\e{{\boldsymbol e}}
\def\p{{\boldsymbol p}}
\def\q{{\boldsymbol q}}
\def\s{{\boldsymbol s}}
\def\bpsi{{\boldsymbol \psi}}
\def\bphi{{\boldsymbol \phi}}

\def\wedgecomma{\mathop{\wedge}\limits_{'}}

\begin{document}
\setcounter{page}{53}

\kwams
{
35Q15, 31A25, 37K10, 35Q58 
}
{
Inverse Scattering Transform, Multi-component NLS equations, Lax representation, The group of reductions
}
\head
{53--67
}
{New types of two component NLS-type equations
}
{V. S. Gerdjikov$^{1, 2}$, A.A. Stefanov$^{1, 3}$\\
 \vspace{0.8cm}
%
}
{A new type of two component NLS-type equations
}
{V. S. Gerdjikov, A.A. Stefanov 
}
{26
}
{2016
}

\begin{abstract}
\noindent{
We study  MNLS related to the D.III-type symmetric spaces. Applying to them Mikhailov reduction groups of the
type $\mathbb{Z}_r\times \mathbb{Z}_2$ we derive new types of 2-component NLS equations. These are {\bf not} counterexamples to
the Zakharov-Schulman theorem because the corresponding interaction Hamiltonians depend not only on $|q_k|^2$, but also on
$q_1q_2^* +q_1^* q_2$.}
\end{abstract}

\section{Introduction}
The non-linear Schrodinger equation
\begin{equation}\label{eq:1}
i \frac{\partial u}{ \partial t} +\frac{1}{2} \frac{\partial ^2 u}{ \partial x^2} +|u|^2u=0,\quad u=u(x,t)
\end{equation}
was first solved by Zakharov and Shabat in 1971 \cite{ZS*71}. Since then, it has found numerous applications \cite{FaTa, ZMNP}.
The first multi-component NLS with applications to physics is the Manakov model \cite{Man*74a, Man*74aa}:
\begin{equation}\label{eq:man}\begin{aligned}
i \frac{\partial u_1}{ \partial t} + \frac{1}{2}\frac{\partial ^2 u_1}{ \partial x^2} +(|u_1|^2 +|u_2|^2)u_1=0, \\
i \frac{\partial u_2}{ \partial t} + \frac{1}{2}\frac{\partial ^2 u_2}{ \partial x^2} +(|u_1|^2 +|u_2|^2)u_2=0.
\end{aligned}\end{equation}
It is natural to look for other types of
multi-component generalizations. Such generalizations were analyzed  in \cite{VNG-6} and this work can be viewed as its continuation.
There is a close relationship between MNLS equations and homogeneous and symmetric spaces \cite{ForKu*83}.
Soon after the pioneer paper by Zakharov and Shabat \cite{ZS*71}, Manakov \cite{Man*74a, Man*74a} proposed a two-component NLS model.
Both NLS models have important applications in nonlinear optics, plasma physics, hydrodynamics etc. Manakov model was easily generalized to
$N$-components known as vector NLS; it has also non-Euclidean version (see \cite{MPK}):
\begin{equation}\label{eq:man2}\begin{aligned}
i \frac{\partial u_1}{ \partial t} + \frac{1}{2}\frac{\partial ^2 u_1}{ \partial x^2} +(|u_1|^2 -|u_2|^2)u_1=0, \\
i \frac{\partial u_2}{ \partial t} + \frac{1}{2}\frac{\partial ^2 u_2}{ \partial x^2} +(|u_1|^2 -|u_2|^2)u_2=0.
\end{aligned}\end{equation}
Here we should mention another famous paper by Zakharov and Schulman \cite{ZS*82} where they prove a theorem classifying the
integrable two-component NLS systems. They request that the interaction Hamiltonian depends only on $|u_1|^2$ and  $|u_2|^2$ and
prove that eqs. (\ref{eq:man}) and (\ref{eq:man2}) are the only integrable MNLS.
 
 The next step in studying multicomponent NLS equations
is based on the important idea of Fordy and Kulish relating the MNLS equations to the symmetric spaces \cite{ForKu*83}.

The present paper is continuation of  \cite{ForKu*83} and a sequel of papers \cite{VSG*94,1,1a, VNG-6,VSG*05,ggk-05,GGK-gik,78,99} in which
Mikhailov's reduction group \cite{Mikhailov} was applied on the MNLS thus deriving new versions with small number of components.
Below we limit ourselves with the MNLS related to the D.III-type symmetric spaces and using Mikhailov reduction groups of the
type $\mathbb{Z}_r\times \mathbb{Z}_2$ derive new types of 2-component NLS equations. These are {\bf not} counterexamples to
the Zakharov-Schulman theorem because the corresponding interaction Hamiltonians depend not only on $|u_k|^2$, but also on
$u_1u_2^* +u_1^* u_2$.

In Section 2 we collect preliminary facts about the D.III symmetric spaces and the types of reductions that
will be applied to the Lax pairs.  In Section 3 we analyze two types of $\mathbb{Z}_4$-reductions of MNLS
related to the algebra ${so}(8)$. In Section 4 we analyze $\mathbb{Z}_5$-reductions of MNLS related to the algebra ${so}(10)$.
To each of these cases we relate a new two-component NLS equation. In Section 5 we formulate the consequences of these
reductions for the Jost solutions and the scattering matrix, We finish with discussions and conclusions.

\section{Preliminaries}

\subsection{Symmetric spaces and $\mathbb{Z}_2$-gradings}
We assume that reader is familiar with the basic properties of the simple Lie groups and Lie algebras, see \cite{Helg}.
The Cartan-Weyl basis of the algebras of the $\rm D_r$-series, $r\geq 4$ are given in the Appendix.

Here will briefly remind the well known facts about the D.III-type symmetric spaces which is $SO^*(2r)/U(r)$
and the structure of its local coordinates \cite{Helg}. Each of these symmetric spaces is generated by a Cartan involution
which induces a $\mathbb{Z}_2$-grading on the Lie algebra which  is $\mathfrak{g}\simeq so(2r)$. The root
system of $\mathfrak{g}$ is:
\begin{equation}\label{eq:Delta}\begin{split}
\Delta = \Delta^+ \cup (-\Delta^+), \qquad \Delta^+ \equiv \{ e_i - e_j, \quad  e_i + e_j, \quad 1 \leq i < j \leq r \},
\end{split}\end{equation}
The $\mathbb{Z}_2$-grading is induced by the Cartan element {red} $J= \sum_{s=1}^{r}H_s$. It induces a $\mathbb{Z}_2$-grading
of  $so(2r)\equiv  \mathfrak{g}^{(0)} \oplus  \mathfrak{g}^{(1)}$ as follows:
\begin{equation}\label{eq:gX}\begin{split}
 \mathfrak{g}^{(0)} \equiv \{ X \in  \mathfrak{g}
\colon [J,X]=0 \}, \qquad  \mathfrak{g}^{(1)} \equiv \{ Y \in  \mathfrak{g} \colon JY+YJ=0 \}.
\end{split}\end{equation}
This grading splits the set of the positive roots $ \Delta^+ = \Delta^+_0 \cup \Delta^+_1$ into subsets:
\begin{equation}\label{eq:Delta1}\begin{aligned}
\Delta^+_0 \equiv \{ e_i - e_j, \quad 1 \leq i < j \leq r \} ,
 \qquad \Delta^+_1 \equiv \{ e_i + e_j, \quad 1 \leq i < j \leq r \}
 \end{aligned}\end{equation}
i.e. $\alpha \in \Delta^+_0$ iff $ \alpha(J) =0$, $\beta \in \Delta^+_1$ iff $\beta(J) =2$.
We will need also the co-adjoint orbit in $\mathfrak{g}$ passing through $J$ which coincides with the linear functionals
(depending on $x$ and $t$) over the linear subspace $\mathfrak{g}^{(1)}$. We will denote it by $\mathcal{M}_J$; a generic $Q(x,t)$ element in it
is provided by:
\begin{equation}\label{eq:Qxt}\begin{split}
 Q(x,t) = \sum_{ \beta \in \Delta_1^+}^{} (q_\beta(x,t) E_\beta + p_\beta(x,t) E_{-\beta} )  = \left(\begin{array}{cc} 0 & \q \\ \p & 0  \end{array}\right).
\end{split}\end{equation}
where $\q(x,t)$ and $\p(x,t)$ are $r\times r$ block matrices.
For simplicity  we will assume that the  $q_\beta(x,t)$ and  $p_\beta(x,t)$ are Schwartz-type functions of their
variables.

\subsection{Reductions}

An important and systematic tool to construct new integrable NLEE is the so-called  reduction group \cite{Mikhailov}.
We will  start with the local $\mathbb{Z}_2$-reductions:
\begin{align}
\label{eq:Z2-Mi1}  A_1U^{\dagger}(x,t,\kappa _1\lambda ^*)A_1^{-1} &= U(x,t,\lambda ),
&\; A_1V^{\dagger}(x,t,\kappa _1\lambda ^*)A_1^{-1} &= V(x,t,\lambda ), \\
\label{eq:Z2-Mi2} A_2U^{T}(x,t,\kappa _2\lambda )A_2^{-1} &= -U(x,t,\lambda ), &\; A_2V^{T}(x,t,\kappa _2\lambda )A_2^{-1} &= -V(x,t,\lambda ), \\
\label{eq:Z2-Mi3} A_3U^{*}(x,t,\kappa _1\lambda ^*)A_3^{-1} &= -U(x,t,\lambda ), &\; A_3V^{*}(x,t,\kappa _1\lambda^* )A_3^{-1} &= -V(x,t,\lambda ), \\
\label{eq:Z2-Mi4} A_4U(x,t,\kappa _2\lambda )A_4^{-1} &= U(x,t,\lambda ), &\; A_4V(x,t,\kappa _2\lambda )A_4^{-1} &= V(x,t,\lambda ).
\end{align}
The consequences of these reductions and the constraints they impose on the FAS and the Gauss
factors of the scattering matrix are well known, see \cite{Mikhailov,ZMNP}.
Since we are dealing with $D_r$-algebras we take into account that $X\to -X^T$ is an inner automorphism.
This means that reduction \eqref{eq:Z2-Mi1} is equivalent to reduction \eqref{eq:Z2-Mi3} and reduction \eqref{eq:Z2-Mi2} is
equivalent to reduction of \eqref{eq:Z2-Mi4}. Therefore it will be enough to consider only reductions \eqref{eq:Z2-Mi3} and \eqref{eq:Z2-Mi4}.

Along with $\mathbb{Z}_2$ we will need also $\mathbb{Z}_p$-reductions, with $p>2$. If $p$ is odd, we can
use only reductions of type \eqref{eq:Z2-Mi4}. The reductions are:
\begin{align}
\label{eq:Zp-Mi1} B_1U^{\dag}(x,t,\kappa_1 (\lambda ^*))B_1^{-1} &= U(x,t,\lambda ), &\; B_1V^{\dag}(x,t,\kappa _1(\lambda^*) )B_1^{-1} &= V(x,t,\lambda ), \\
\label{eq:Zp-Mi2} B_4U(x,t,\kappa _4(\lambda ))B_4^{-1} &= U(x,t,\lambda ), &\; B_4V(x,t,\kappa _4(\lambda ))B_4^{-1} &= V(x,t,\lambda ).
\end{align}
where the functions $\kappa_3$ and $\kappa_4$ if applied $p$ times satisfy
$\kappa_j(\kappa_j(...\kappa_j(\lambda)...)) =\lambda$, $j=3,4$.
In addition we will use also $\mathbb{Z}_r$-reductions of the form
\begin{equation}\label{eq:Zp}\begin{aligned}
 A_1U(x,t,  \epsilon \lambda )A_1^{-1} = U(x,t,\lambda ), \quad A_1V(x,t,\epsilon\lambda )A_1^{-1} = V(x,t,\lambda )
\end{aligned}\end{equation}
where $\epsilon =\pm 1$ and $A_1^r =\openone $ with $r>2$; if $r$ is odd then $\epsilon =1$.  We will demonstrate below
examples when such reductions with $\epsilon=1$ provide new types of NLEE.

\subsection{Lax pair and reductions}
Below we outline the formulation of the Lax pair for the D.III-type symmetric spaces:
\begin{equation}\begin{split}\label{eq:Mop}
&L(\lambda)\psi \equiv \left(i{d\over dx} + Q(x,t) -  \lambda J\right)  \psi (x,t,\lambda )=0,\\
 &M(\lambda)\psi \equiv \left(i{d\over dt} - \frac{1}{2}[Q, \ad_J^{-1}Q] + i  \mbox{ad}_J^{-1} \frac{\partial Q}{ \partial x } +\lambda Q  - \lambda ^2 J \right)
 \psi (x,t,\lambda )=0,
 \end{split}\end{equation}
where $Q=Q(x,t)$ and $J$ are elements of the algebra $so(2r)$. In other words they are $2r \times 2r$ matrices with the following
block-matrix form:
\begin{equation}\label{eq:QJ}\begin{aligned}
 Q(x,t) = \left(\begin{array}{cc} 0 & \q(x,t) \\ \p(x,t) & 0   \end{array}\right), \qquad J = \left(\begin{array}{cc}\openone & 0 \\
0 & -\openone  \end{array}\right).
\end{aligned}\end{equation}

 The compatibility condition $[L(\lambda),M(\lambda)]=0$ of the operators in  (\ref{eq:Mop}) gives the general form
of the D.III-type MNLS equations on symmetric spaces. It can be viewed as block-matrix generalization of the AKNS system \cite{AKNS};
see also \cite{APT}:
\begin{equation}\label{eq:MNLSeq}
{i\over 2} \left[J,{dQ \over dt }\right] + \frac{1}{2} {d^2Q  \over dx^2 } - \left[\ad_J^{-1}Q,[\ad_J^{-1}Q,Q] \right]=0.
\end{equation}

Consider the Lax pair of the Zakharov - Shabat system
\begin{equation}
\begin{aligned}
L \psi(x,t, \lambda) &=\left( i \frac{\partial }{ \partial x } + U(x,t,\lambda) \right) \psi(x,t, \lambda)=0, &\quad U &=  Q(x,t) - \lambda J,  \\
M \psi(x,t, \lambda) &=\left( i \frac{\partial }{ \partial t} + V(x,t,\lambda)\right) \psi(x,t, \lambda)=0,  &\quad V &= V_0 + \lambda Q - \lambda^2 J.
\end{aligned}
\label{LaxPair}
\end{equation}

Let $\mathcal{M}_J $ be the co-adjoint orbit  of $\mathfrak{g} $ passing
through $J$. Then $Q(x,t) \in \mathcal{M}_J $.

As mentioned above, the choice of $J $ determines the dimension of $\mathcal{M}_J $ which
can be viewed as the phase space of the relevant nonlinear evolution
equations (NLEE).  It is equal to the number of roots of $ \mathfrak{g}
$ such that $\alpha (J)\neq 0 $.  Taking into account that if $\alpha  $
is a root, then and $-\alpha $ is also a root of $\mathfrak{g} $ then $\dim \mathcal{M}_J $ is always even.
Since  all the examples are related to symmetric spaces of D.III-type it is natural to choose
$J $ as in (\ref{eq:QJ}).
As a consequence $\mathfrak{g}^{(0)}$ ( which can be viewed as the kernel of the  operator $\ad_J $) is non-commutative
and isomorphic to $so(r)\oplus so(r)$.

Below the automorphisms $A_i$ and $B_k$ used for reductions will be inner and will correspond to compositions of Weyl reflections.
They will act as similarity transformations with $2r\times 2r$ matrices  that belong to the $SO(2r)$ group and satisfy
\begin{equation}
A^r (X) A^{-r} \equiv  X , \qquad  B^r (X) B^{-r} \equiv  X , \qquad \forall X \in \mathfrak{g}.
\end{equation}
From eqs. (\ref{eq:Zp}) and (\ref{LaxPair}) there follows that they must either preserve $J$ or change its sign, i.e.
\begin{equation}\label{eq:AB0}\begin{aligned}
& AJA^{-1}  = J, &\quad  &BJB^{-1}  = -J, \\
&A = \begin{pmatrix} \a_1 & 0 \\ 0 & -\a_1^T \end{pmatrix}, &\quad &B = \begin{pmatrix} 0 & \b_1 \\ \b_2 & 0 \end{pmatrix},  \\
& \hat{\a}_2 = \hat{\s}_1 \a_1^T \s_1, &\quad  & \hat{\b}_2 = \s_1 \b_1^T \s_1 .
\end{aligned}\end{equation}


The compatibility condition $ [L,M]=0$ leads to the MNLS of the form:
\begin{equation}
i\frac{\partial Q}{ \partial t } + \frac{1}{2} \frac{\partial^2 Q }{ \partial x^2 } +  [Q, V_0]=0.
\label{EqGen}
\end{equation}

We will consider further reductions of the above system. Namely, we will consider reduction of type $\mathbb{Z}_r$ of the form
\begin{equation}\label{Reduction1}\begin{aligned}
A U(x,t,\lambda) A^{-1} &= U(x,t,\lambda), &\quad  A V(x,t, \lambda) A^{-1} &= V(x,t, \lambda), \\
B U(x,t,-\lambda) B^{-1} &= U(x,t,\lambda), &\quad  B V(x,t, -\lambda) B^{-1} &= V(x,t, \lambda),
\end{aligned}\end{equation}
where $A$ and $B$  are automorphisms of order $r$.
This will restrict the number of the independent variables in $Q$ to twice the number of orbits of $A$.  This means that the potential of $L$ takes the form
\begin{equation}\label{eq:QAB}\begin{aligned}
Q_A &= \sum_{\alpha \in \delta_A^+} q_{\alpha}(x,t)  \mathcal{E}_{\alpha}^{A} + p_{\alpha}(x,t) \mathcal{E}_{-\alpha}^{A}, &\quad
\mathcal{E}_{\alpha}^{A} &= \sum_{ s=0}^{r-1} A^{s} E_\alpha A^{-s},  \\
Q_B &= \sum_{\alpha \in \delta_B^+} q_{\alpha}(x,t)  \mathcal{E}_{\alpha}^{B} + p_{\alpha}(x,t) \mathcal{E}_{-\alpha}^{B}, &\quad
\mathcal{E}_{\alpha}^{B} &= \sum_{ s=0}^{r-1} B^{s} E_\alpha B^{-s},
\end{aligned}\end{equation}
and $\delta_A^+$ and $\delta_B^+$   contains  only one root $\alpha$ from each orbit of the corresponding automorphism.
In addition we will impose also a  $\mathbb{Z}_2$ reductions of the type
\begin{equation}
\label{Reduction2}
\begin{aligned}
U^{\dagger}(x,t,\lambda^*) = U(x,t,\lambda), \qquad V^{\dagger}(x,t,\lambda^*) = V(x,t,\lambda).
\end{aligned}
\end{equation}
This reduction will restrict the form of $Q$ to $p = q^{\dagger}$.

%
\section{MNLS with $\mathbb{Z}_4$-reductions related to $D_4$}
We will consider $\mathfrak{g} = D_4$ and $J = \mbox{diag}(1,1,1,1,-1,-1,-1,-1)$. There are two realizations of the reduction \eqref{Reduction1} that will give two-component MNLS:
\begin{equation}\label{eq:AB1}\begin{split}
A = S_{e_1- e2} \circ S_{e_2 - e_3}\circ S_{e_3 - e_4} , \qquad B = S_{e_1+ e_2} \circ S_{e_2 + e_3}\circ S_{e_3 + e_4}
\end{split}\end{equation}
In the first case  $A$ acts in the root space by $A: e_1 \to e_2 \to e_3 \to e_4 \to e_1$
and splits $\Delta$ into 8 orbits
\begin{equation}
\begin{aligned}
&\mathcal{O}_1^\pm : \pm ( e_1 +e_2)\to \pm (e_2 + e_3) \to  \pm (e_3 + e_4) \to \pm (e_1 + e_4), \\
&\mathcal{O}_2^\pm : \pm (e_1 + e_3) \to  \pm (e_2 + e_4), \\
&\mathcal{O}_3^\pm : \pm (e_1 - e_2) \to \pm (e_2 - e_3) \to \pm (e_3 - e_4) \to \pm (e_4 - e_1), \\
&\mathcal{O}_4^\pm : \pm (e_1 - e_3) \to \pm (e_2 - e_4) \to \pm (e_3 - e_1) \to \pm (e_4 - e_2).
\end{aligned}
\end{equation}

four of which $\mathcal{O}_1^\pm \cup \mathcal{O}_2^\pm$ span the set of roots $\Delta_1^+\cup \Delta_1^-$; the other
four orbits $\mathcal{O}_3^\pm \cup \mathcal{O}_4^\pm$ span the set of roots $\Delta_0^+\cup \Delta_0^-$.
This $\mathbb{Z}_4$ reduction is realized by a type-A automorphism as in (\ref{eq:AB0}) with
\begin{equation}
\a_1 =
\begin{pmatrix}
0 & 0 & 0 & -1 \\
1 & 0 & 0 & 0 \\
0 & -1 & 0 & 0 \\
0 & 0 & 1 & 0
\end{pmatrix}.
\end{equation}

In the second case  $B$ acts in the root space by $B: e_1 \to - e_2 \to e_3 \to - e_4 \to e_1$
and splits $\Delta$ into 8 orbits
\begin{equation}
\begin{aligned}
& \tilde{ \mathcal{O}}_1^\pm : \pm ( e_1 +e_2)\to \mp (e_2 + e_3) \to  \pm (e_3 + e_4) \to \mp (e_1 + e_4), \\
&\tilde{ \mathcal{O}}_2^\pm : \pm (e_1 + e_3) \to  \mp (e_2 + e_4), \\
&\tilde{ \mathcal{O}}_3^\pm : \pm (e_1 - e_2) \to \mp (e_2 - e_3) \to \pm (e_3 - e_4) \to \mp (e_4 - e_1), \\
&\tilde{ \mathcal{O}}_4^\pm : \pm (e_1 - e_3) \to \mp (e_2 - e_4) \to \pm (e_3 - e_1) \to \mp (e_4 - e_2).
\end{aligned}
\end{equation}
four of which $\tilde{ \mathcal{O}}_1^\pm \cup \tilde{ \mathcal{O}}_2^\pm$ span the set of roots $\Delta_1^+\cup \Delta_1^-$; the other
four orbits $\tilde{ \mathcal{O}}_3^\pm \cup \tilde{ \mathcal{O}}_4^\pm$ span the set of roots $\Delta_0^+\cup \Delta_0^-$.
%
We will consider the first case. The potential of the Lax operator  $Q(x,t)$ (see \eqref{eq:Qxt}) is given by
\begin{equation}
\q (x,t) = \begin{pmatrix} q_1 & \sqrt{2} q_2 & q_1 & 0 \\
\sqrt{2} q_2 & q_1 & 0 & q_1 \\ -q_1 & 0 & q_1 & -\sqrt{2} q_2 \\
 0 & -q_1 & -\sqrt{2} q_2 & q_1 \end{pmatrix}, \qquad
\p(x,t) = \q^\dag (x,t),
\end{equation}

Imposing also the second reduction \eqref{Reduction2} ($p_i = q_i^{*}$) the equations become
\begin{equation}
\begin{aligned}
 i \frac{\partial q_1}{ \partial t }+  \frac{1}{2} \frac{\partial q_1}{ \partial x^2 }  + 2 q_1 ( |q_1|^2 + 2 |q_2|^2) + 2  q_2^2 q_1^* &=0, \\
 i \frac{\partial q_2}{ \partial t }+  \frac{1}{2} \frac{\partial q_2}{ \partial x^2 } + 2 q_2 (2 |q_1|^2 + |q_2|^2) + 2 q_1^2 q_2^{*} &=0.
\end{aligned}
\end{equation}
They  admit a Hamiltonian formulation, with a Hamiltonian density given by
\begin{equation}
\mathcal{H} = \frac{1}{2} \left| \frac{\partial q_1}{ \partial x }\right|^2 + \frac{1}{2} \left| \frac{\partial q_2}{ \partial x }\right|^2
 - (|q_1|^2 +  |q_2|^2)^2 - (q_1 q_2^* + q_1^* q_2)^2
\end{equation}

\section{MNLS with $\mathbb{Z}_5$-reductions related to $D_5$}

Let  $\mathfrak{g} = D_4$ and $J = \mbox{diag}(1,1,1,1,1,-1,-1,-1,-1,-1)$. Again, we will also impose \eqref{Reduction1}
with $A = S_{e_1- e2} \circ S_{e_2 - e_3}\circ S_{e_3 - e_4} \circ S_{e_4 - e_5}$, that is $A: e_1 \to e_2 \to e_3 \to e_4 \to e_5$.
This splits $\Delta$ into orbits
\begin{equation}
\begin{aligned}
&\mathcal{O}_1^\pm : \pm (e_1 +e_2) \to  \pm (e_2 + e_3) \to  \pm (e_3 + e_4) \to  \pm (e_4 + e_5) \to  \pm (e_1 + e_5), \\
&\mathcal{O}_2^\pm: \pm (e_1 + e_3) \to  \pm (e_2 + e_4) \to  \pm (e_3 + e_5) \to  \pm (e_1 + e_4) \to  \pm (e_2 + e_5), \\
&\mathcal{O}_3^\pm:  \pm (e_1 - e_2) \to \pm (e_2 - e_3) \to  \pm (e_3 - e_4) \to  \pm (e_4 - e_5) \to \mp (e_1 - e_5) , \\
&\mathcal{O}_4^\pm: \pm (e_1 - e_3) \to \pm (e_2 - e_4) \to  \pm (e_3 - e_5) \to \mp (e_1 - e_4) \to \mp(e_2 - e_5).
\end{aligned}
\end{equation}
This $\mathbb{Z}_5$ reduction is realized by a type-A automorphism as in (\ref{eq:AB0}) with
\begin{equation}
\a_1 = \begin{pmatrix} 0 & 0 & 0 & 0 & 1 \\ 1 & 0 & 0 & 0 & 0 \\ 0 & -1 & 0 & 0 & 0 \\
0 & 0 & 1 & 0 & 0 \\0 & 0 & 0 & -1 & 0 \end{pmatrix}.
\end{equation}

The reduced potential $Q(x,t)$ as in (\ref{eq:QJ}) with
\begin{equation}
\q(x,t) = \begin{pmatrix} q_1 & q_2 & q_2 & q_1 & 0 \\ q_2 & -q_2 & q_1 & 0 & q_1 \\ q_2 & q_1 & 0 & q_1 & -q_2 \\
q_1 & 0 & q_1 & q_2 & q_2 \\ 0 & q_1 & -q_2 & q_2 & -q_1 \end{pmatrix}, \quad
\p(x,t) =\begin{pmatrix} p_1 & p_2 & p_2 & p_1 & 0 \\ p_2 & -p_2 & p_1 & 0 & p_1 \\ p_2 & p_1 & 0 & p_1 & -p_2 \\
p_1 & 0 & p_1 & p_2 & p_2 \\ 0 & p_1 & -p_2 & p_2 & -p_1 \end{pmatrix}.
\end{equation}
 After  imposing the second reduction  $\p = \q^\dag$, or $p_i = q_i^*$ the equations become
\begin{equation}\begin{aligned}
 i \frac{\partial q_1}{ \partial t }  +  \frac{1}{2} \frac{\partial^2 q_1}{ \partial x^2}  +
  q_1(3 |q_1|^2  + 4  |q_2|^2) +    q_2(2 |q_1|^2 -   |q_2|^2) + q_1^2 q_2^* +2 q_1^* q_2^2  &=0, \\
 i \frac{\partial q_2}{ \partial t }  +  \frac{1}{2} \frac{\partial^2 q_2}{ \partial x^2}  +  q_1( |q_1|^2 -2 |q_2|^2)
 + q_2 (4 |q_1|^2 + 3 |q_2|^2)  - q_1^* q_2^2  +2  q_1^2 q_2^*   &=0.
\end{aligned}
\end{equation}
The above equations admit a Hamiltonian formulation, with a Hamiltonian density given by
\begin{equation}
\begin{aligned}
\mathcal{H} &= \frac{1}{2} \left| \frac{\partial q_1}{ \partial x }\right|^2 + \frac{1}{2} \left| \frac{\partial q_2}{ \partial x }\right|^2 - (|q_1|^2 + |q_2|^2 )^2 \\
 &- \frac{1}{2} ( |q_1|^2 + q_1^* q_2 + q_1 q_2^*)^2 - \frac{1}{2} ( |q_2|^2 - q_1^* q_2 - q_1 q_2^*)^2
\end{aligned}
\end{equation}

\section{Direct and inverse scattering problems}

Basic tools in this analysis are the Jost solutions (we will avoid writing explicit time dependence, to avoid cluttering the notation)
\begin{equation}\label{eq:Jo}\begin{split}
\lim_{x\to -\infty}\phi (x,\lambda) e^{iJ\lambda x} =\openone, \qquad \lim_{x\to \infty}\psi (x,\lambda) e^{iJ\lambda x} =\openone.
\end{split}\end{equation}
Formally the Jost solutions must satisfy Volterra type integral equations.
If we introduce
\begin{equation}
\label{eq:xipm}
\xi (x,\lambda) =\psi (x,\lambda) e^{i\lambda Jx}, \quad \eta (x,\lambda) =\phi (x,\lambda) e^{i\lambda Jx},
\end{equation}
then $\xi_\pm (x,\lambda)$ must satisfy
\begin{equation}
\label{eq:xipm'}
\begin{split}
\xi(x,t) &= \openone + i \int_{\infty}^{x} dy \; e^{-i\lambda J(x-y)} Q(y,t) \xi (y,\lambda)  e^{i\lambda J(x-y)}, \\
\eta(x,t) &= \openone + i \int_{-\infty}^{x} dy \; e^{-i\lambda J(x-y)} Q(y,t) \eta (y,\lambda)  e^{i\lambda J(x-y)}. \\
\end{split}
\end{equation}
The Jost solutions can not be extended for $ \mbox{Im} \lambda \neq 0$. However some of their columns can be extended for
$\lambda \in \mathbb{C}_+$ or $\lambda \in \mathbb{C}_-$. The Jost solutions can be written in the following block-matrix form
\begin{equation}
\psi (x,\lambda) = \begin{pmatrix} \bpsi _1^-(x,\lambda) & \bpsi_1^+ (x,\lambda) \\ \bpsi_2^- (x,\lambda) & \bpsi_2^+ (x,\lambda)\end{pmatrix}
, \quad \phi (x, \lambda) = \begin{pmatrix} \bphi_1^+ (x, \lambda) & \bphi_1^- (x, \lambda) \\ \bphi _2^+(x, \lambda) & \bphi_2^- (x, \lambda) \end{pmatrix},
\end{equation}
where the superscript $\pm$ shows that the corresponding $r \times r$ block allows analytic extension for  $\lambda \in \mathbb{C}_\pm$.
Then the scattering matrix is introduced by
\begin{equation}\label{eq:T}\begin{split}
T(\lambda) = \hat{\psi} ( x, \lambda) \phi ( x,\lambda), \quad T(\lambda) = \begin{pmatrix} \a^+(\lambda) & -\b^-(\lambda) \\
\b^+ (\lambda) & \a^-(\lambda) \end{pmatrix}
\end{split}\end{equation}
where by "hat" we denote matrix inverse. Since without loss of generality the Jost solutions are group elements the scattering matrix will also be a group element.
Note that if we impose one of the $\mathbb{Z}_r$ reduction ($A$ or $B$) on the Lax pair then
$T(\lambda,t)$ must satisfy one of the relations below.
\begin{equation}
A T(\lambda,t) A^{-1} = T(\lambda), \qquad B T(-\lambda,t) B^{-1} = T(\lambda).
\end{equation}
This imposes the following constraints on the blocks $\a^\pm (\lambda)$ and $\b^\pm(\lambda,t)$:
\begin{equation}\label{eq:ATA}\begin{aligned}
 \a^+(\lambda) &= \a_1 \a^+(\lambda) \hat{\a}_1, &\quad \b^-(\lambda,t) &= \a_1 \b^-(\lambda,t) \hat{\a}_2,  \\
 \b^+(\lambda,t) &= \a_2 \b^+(\lambda,t) \hat{\a}_1, &\quad    \a^-(\lambda) &= \a_2 \a^-(\lambda) \hat{\a}_2,
\end{aligned}\end{equation}
for the type-A reductions and
\begin{equation}\label{eq:BTB}\begin{aligned}
 \a^+(\lambda) &= \b_1 \a^-(-\lambda) \hat{\b}_1, &\quad \b^-(\lambda,t) &= -\b_1 \b^+(-\lambda,t) \hat{\b}_2,  \\
 \b^+(\lambda,t) &= -\b_2 \b^-(-\lambda,t) \hat{\b}_1, &\quad    \a^-(\lambda) &= \b_2 \a^+(-\lambda) \hat{\b}_2.
\end{aligned}\end{equation}
where the constant matrices $\a_k$ and  $\b_k$, $k=1,2$ are given in eq. (\ref{eq:AB0}).

The second reduction $Q(x,t)=Q^\dag (x,t)$ imposes on the Jost solutions and on the scattering matrix
the constraints (below ''hat'' denotes matrix inverse) :
\begin{equation}\label{eq:Qdag}\begin{aligned}
\psi^\dag (x,t, \lambda^*) & = \hat{\psi} (x,t,\lambda), &\quad \phi^\dag (x,t, \lambda^*) & = \hat{\phi} (x,t,\lambda), \\
T^\dag (\lambda^* ,t) &= \hat{T}(\lambda ,t), &\quad \hat{T}(\lambda ,t) &= \left(\begin{array}{cc} \c^- & \d^- \\ -\d^+ & \c^+  \end{array}\right).
\end{aligned}\end{equation}
The corresponding blocks of $T(\lambda,t)$ and its inverse $\hat{T}(\lambda,t)$ must satisfy
\begin{equation}\label{eq:Tc}\begin{aligned}
( \a^+)^\dag (\lambda^*) &= \c^-(\lambda), &\quad ( \b^-)^\dag (\lambda^*,t) &= \d^-(\lambda,t), \\
( \b^+)^\dag (\lambda^*,t) &= -\d^+(\lambda,t), &\quad  ( \a^-)^\dag (\lambda^*) &= \c^+(\lambda).
\end{aligned}\end{equation}

We end this section by formulating the time-dependence of the scattering matrix which follows naturally from
the Lax representation (\ref{eq:Mop}):
\begin{equation}\label{eq:Tt}\begin{aligned}
i \frac{\partial T}{ \partial t } - \lambda^2 [J, T(\lambda,t)]=0, \quad 
\mbox {i.e.} \quad  \begin{aligned}
\frac{\partial \a^\pm}{ \partial t} &=0 , &\quad  i \frac{\partial \b^\pm}{ \partial t } \mp b^\pm (\lambda,t) =0, \\
\frac{\partial \c^\pm}{ \partial t} &=0 , &\quad  i \frac{\partial \d^\pm}{ \partial t } \pm b^\pm (\lambda,t) =0.
\end{aligned}
\end{aligned}\end{equation}
In particular, the diagonal blocks can be viewed as generating functionals of the integrals of motion of the MNLS.

\section{Discussion and conclusions}

The inverse scattering method, applied to the scalar NLS equation
has all the properties of a Generalized Fourier Transform (GFT). The derivation of these properties
is based on the Wronskian relations \cite{AKNS,CaDe,CaDe2}. These results allow natural generalizations
to the   MNLS equations, see \cite{VSG*05} and references therein.

The mapping $\mathfrak{F}\colon \mathcal{M}_J \to T(\lambda,t)$ is directly related to the  GFT  which instead of the
usual exponentials $e^{\pm i\lambda J}$ uses the so-called `squared solutions' of $L$, $\e_\alpha^\pm (x,\lambda) = \pi_{0,J} (\chi^\pm E_\alpha
\hat{\chi}^\pm (x,\lambda)$. Here $\chi^\pm(x,\lambda)$ are the FAS of $L$ and $\pi_{0,J} X = \ad_J^{-1} \ad_J X$ is a projector onto the image of the
operator $\ad_J$. Next, one can: i) prove that the system of `squared solution' $\e_\alpha^\pm (x,\lambda) , \alpha \in \Delta_1$ are complete set of
functions on $\mathcal{M}_J$; ii) the minimal set of scattering data can be viewed as expansion coefficients of $Q(x,t)$ over $\e_\alpha^\pm (x,\lambda) $;
ii) the variations of the minimal set of scattering data can be viewed as expansion coefficients of $\ad_J^{-1} \delta Q(x,t)$ over $\e_\alpha^\pm (x,\lambda) $.
Finally these expansions can be used to derive the fundamental properties of the whole class of multi-component NLEE, for more details and proofs see.
\cite{VSG*05}. In particular this means that the MNLS equations  \eqref{EqGen} admit a hierarchy of  hamiltonian formulations,
see \cite{VSG*05}. The simplest of them has as Hamiltonian
\begin{equation}\label{eq:H0}\begin{split}
H^{(0)}&= \frac{1}{2r} \int_{{-\infty}}^{\infty} dx\; \left( \left\langle \frac{\partial Q}{ \partial x }, \frac{\partial Q}{ \partial x } \right\rangle
-  \left\langle \left[ \ad_J^{-1} Q, Q\right], \left[ \ad_J^{-1} Q, Q\right]\right\rangle \right) .
\end{split}\end{equation}
The relevant  symplectic form is given by
\begin{equation}\label{eq:Ome}\begin{split}
\Omega^{(0)} &= \frac{i}{2r} \int_{x=-\infty}^{\infty}dx\;  \left \langle \ad_J^{-1} \delta Q \wedgecomma \ad_J^{-1} \delta Q \right\rangle .
\end{split}\end{equation}
The other members of the hierarchy are generated by the recursion operators $\Lambda_\pm$. The proof of their compatibility
is based on the completeness relations for the `squared solutions' that are eigenfunctions of $\Lambda_\pm$. Thus it is
natural to expect that the new 2-component NLS will also possess hierarchies of Hamiltonian structures.
These type of results can be viewed also as more strict proofs, that have been formally derived by Dickey and Gelfand
\cite{grah:wildgelf,grah:wildgelf2}, by Drinfeld and Sokolov \cite{DrSok} and by Lombardo and Mikhailov
\cite{MiLo1,MiLo2}, see also \cite{78,99}.

Our last remark is that one can apply similar ideas also to the other types of symmetric spaces. Thus
one can find  other 2-component MNLS related to the other symmetric spaces:  BD.I, C.I etc.
These and other natural problems, such as deriving their soliton solutions,
the construction of their integrals of motion, derivation of the fundamental properties etc.
will be published elsewhere.


\section*{Acknowledgements}
One of us (VSG) is grateful to professor F. Calogero for the useful discussion and suggestions.
\section*{Appendix}
\appendix

The simple Lie algebras $D_r \equiv \mathfrak{so}(2r)$ are usually represented by a $2r \times 2r$ antisymmetric matrices.
In this realization the Cartan subalgebra is not diagonal, so we will use a realization for which every $X \in D_r$ satisfies
\begin{equation}
SX+X^TS=0, \qquad  S = \begin{pmatrix} 0 & \s \\   \hat{\s} & 0 \end{pmatrix},  \qquad S^2 = \openone.
\end{equation}
Here the block $\s$ is a $r \times r$ matrix given by $\s = \sum_{k=1}^{r}(-1)^{k+1} E_{k,r+1-k} $, where
$E_{kj}$ are $r\times r$ matrix given by $(E_{kj})_{mn}=\delta_{km}\delta_{jn}$.
With this definition  the Cartan subalgebra is given by diagonal matrices.
The root system $\Delta$ of $D_{r}$ consists of positive roots $\Delta_{+}$ and negative roots $\Delta_{-}$. If $\alpha \in \Delta_{+}$
then $-\alpha \in \Delta_{-}$. If $e_i$ is an orthonormal basis in $\mathbb{R}^r$ then the set of positive roots $\Delta_{+}$ consists of $e_i - e_j$, $e_i + e_j$ with $1 \leq  i <  j  \leq r$.
The Cartan-Weyl basis of $D_r$ is given by
\begin{equation}
\begin{aligned}
&H_i = E_{ii} - E_{2r+1-i,2r+1-i},  &\quad 1 \leq &i \leq r, \\
&E_{e_i - e_j} = E_{ij} - (-1)^{i+j} E_{2r+1-j,2r+1-i}, &\quad  1 \leq  i < & j  \leq r,  \\
&E_{e_i + e_j} = E_{i,2r+1-j} +  (-1)^{i+j} E_{j,2r+1-i} , &\quad  1 \leq  i <  &j  \leq r, \quad E_{-\alpha}=(E_{\alpha})^T,
\end{aligned}
\end{equation}
where by $E_{ij}$ we denote a $2r\times 2r$ matrix that has a one at the $i-th$ row and $j-th$ column and is zero everywhere else.


%
\begin{flushleft}
{\it
V. S. Gerdjikov \\
 Institute of Mathematics and Informatics \\
Bulgarian Academy of Sciences \\
Acad. Georgi Bonchev Str., Bl. 8 \\
1113 Sofia, Bulgaria \\
and \\
 Institute for Nuclear Research and Nuclear Energy \\
Bulgarian Academy of Sciences \\
72 Tzarigradsko chausee\\
Sofia 1784, Bulgaria \\
e-mail: vgerdjikov@math.bas.bg \\
\vspace{0.3cm}

%
 A.A. Stefanov \\
Institute of Mathematics and Informatics, \\
Bulgarian Academy of Sciences, \\
Acad. Georgi Bonchev Str., Bl. 8 \\
1113 Sofia, Bulgaria, \\
and \\
 Faculty of Mathematics and Informatics \\
Sofia University "St. Kliment Ohridski" \\
5 James Bourchier Blvd \\
 1164 Sofia, Bulgaria \\
e-mail: aleksander.a.stefanov@gmail.com
}
\end{flushleft}

\end{document}